\documentclass[12pt]{article}

\usepackage[utf8]{inputenc}
\usepackage{color}
\usepackage{mathtools}
\usepackage{amssymb}
\usepackage{graphics}
\graphicspath{./fig/}
\usepackage{authblk}
\usepackage{subcaption}
\usepackage{delarray}
\usepackage{setspace}
\usepackage{bm}
\usepackage{float}
\usepackage{newtxtext,newtxmath}  
\usepackage[sort&compress,round,colon,authoryear]{natbib}
\usepackage{multirow}

\usepackage{color,soul}

\usepackage[margin=1in]{geometry}

\newcommand{\acronym}{RATEC}

\begin{document}
\title{A Multi-Channel Approach for Automatic Microseismic Event Association using {RANSAC}-based Arrival Time Event Clustering (\acronym)}
\author[1]{Lijun Zhu\thanks{lijun.zhu@gatech.edu}}
\author[2]{Lindsay Chuang}
\author[1]{James H. McClellan}
\author[1]{Entao Liu}
\author[2]{Zhigang Peng}
\affil[1]{ECE at Georgia Institute of Technology}
\affil[2]{EAS at Georgia Institute of Technology}

\renewcommand\Authands{, and }

\maketitle
\newpage
\thispagestyle{empty}

\listoffigures
\listoftables
\newpage

\begin{abstract}
  In the presence of background noise, arrival times picked from a surface microseismic data set usually include a number of false picks that can lead to uncertainty in location estimation.
  To eliminate false picks and improve the accuracy of location estimates, we develop an association algorithm termed RANSAC-based Arrival Time Event Clustering (\acronym) that clusters picked arrival times into event groups based on random sampling and fitting moveout curves that approximate hyperbolas.
  Arrival times far from the fitted hyperbolas are classified as false picks and removed from the data set prior to location estimation.
  Simulations of synthetic data for a 1-D linear array show that \acronym\ is robust under different noise conditions and generally applicable to various types of subsurface structures.
  By generalizing the underlying moveout model, \acronym\ is extended to the case of a 2-D surface monitoring array.
  The effectiveness of event location for the 2-D case is demonstrated using a data set collected by the  5200-element dense Long Beach array. The obtained results suggest that \acronym\ is effective in removing false picks and hence can be used for phase association before location estimates.

\end{abstract}
\newpage


\newpage

\section{Introduction}

\label{sec:introduction}

Microseismic event monitoring is an important tool in understanding the dynamic state of subsurface structure during hydraulic fracturing operation.
The source mechanisms and locations of these microearthquakes can help characterize geometries and spatio-temporal evolution of fractures generated during fluid injections.
Microseismic events are typically at low magnitudes and occur closely in space and time, presenting significant challenges for phase picking, association and event location (\citealt{eaton2018passive}).
After initial phase picking, locating earthquakes involves two folds of operations: phase association and location estimation. The former refers to the procedure where a set of seismic phases are grouped into a common source, and the latter refers to solving for earthquake locations given a set of phases. Although less attention had been put on phase association in the past, it is now becoming a significant problem due to the recent improvement of detection abilities on weak seismic phases, mostly with deep-learning techniques (\citealt{Ross2018}; \citealt{zhu2019deep}; \citealt{Mousavi2020}).

In global seismology, the association problem has been formulated as variants of statistical problems and solved by grid-search (\citealt{Lebras1994};\citealt{Rodi2000};\citealt{Draelos2015}). These algorithms search for the best grid that can maximize the coherency of observed and predicted arrival times. Formulation of such type can be categorized as the back-projection method (\citealt{KiserIshii2017}). At regional and local distances, sophisticated workflows based on back-projection or a grid-search algorithm have been created in order to meet different monitoring needs, e.g., Earthworm, SeisCompP3, GLASS3, REAL, (\citealt{Yeck2019}; \citealt{Zhang2019}).
More recently \cite{Ross2019} developed a deep-learning-based framework to solve association problems via modeling the temporal and contextual relationships of seismic phases.
\cite{McBrearty2019} utilized graph theory to solve for phase association and source location simultaneously for area-specific back-projection problems.
A reliable phase association result is the foundation of an accurate event location estimate.

Various earthquake location methods are implemented as standard procedures in data process pipelines.
In earthquake seismology, the common practice is to use geometric methods (i.e., triangulation) based upon direct arrivals picked at individual receivers (\citealt{Zhang2003}).
Such a method is simple and only requires low computational cost, but it is only viable for borehole data (\citealt{Maxwell2010}) as arrivals detected on traces from surface data are less reliable when signal-to-noise ratios (SNR) are low (\citealt{DunEis2010}).
Alternatively, semblance-based methods have been developed that do not rely on picked arrival times.
These methods utilize semblance measures that involve stacking waveforms according to forward calculated travel times (\citealt{Duncan2005}; \citealt{Lakings2006};\citealt{Tan2014}; \citealt{Frantisek2015}).
In order to find the optimal location which yields the best stacking results, these methods have to discretize a monitoring region and then perform a grid search, which demands high computational resources.
Although attempts have been made to accelerate the exhaustive search for the best location by global optimization such as differential evolution (\citealt*{Zhu2015}) and particle swarm (\citealt*{Luu2016}), these methods are, in general, slow on large monitoring regions with high spatial resolution requirements.
Another set of methods use reverse time migration (RTM) to find event locations (\citealt{GajTes2005}; \citealt{Artman2010}; \citealt{NakBer2016}). These methods take advantage of waveform redundancies existing among receivers, and backward propagate wavefields to the source location in space and time through assumed models. The optimal locations are defined at the points where the energies are most focused.
Recent development of full-waveform inversion (FWI) has inspired full-wave based methods (\citealt{WittenShragge2016}; \citealt{Sharan2016}). However, like the RTM based methods, the finite-difference (FD) modeling they rely on is computationally intensive and can be slower than the grid search program in travel-time based methods.

The recent development of hydraulic fracturing produces increasing amounts of data from passive monitoring, which in turn demands a more efficient processing scheme for surface arrays that can be deployed without drilling monitoring wells.
A recent study by \cite{AkramEaton2016} summarized and compared the most common arrival time picking methods on a single trace, such as short-term over long-term average ratio (STA/LTA) (\citealt{Allen1978}; \citealt{EarShe1994}), a modified energy ratio (MER) (\citealt*{Han2009}), a modified form of Coppens' method (MCM) (\citealt{SabVel2010}), and Akaike's information criterion (AIC) (\citealt{Tak1991}), to name a few.
A common theme in all of these methods is the use of processing to increase the significance of detected peaks and minimize the number of false picks from noisy data recorded by surface arrays, which leads to bad event location estimation.
Ideally, when all the picks are perfect, a moveout curve can be computed to fit exactly through the arrival-time picks. Alternatively, \cite*{Zhu2016} took advantage of the fact that a group of receivers with both good and bad picks still \emph{contains a subset of picks that follow an expected trend} of arrival times when events are present.
By fitting a moveout curve through subsets of picked arrival times using random sample consensus (RANSAC) (\citealt{FisBol1981}), \cite{Zhu2016} were able to recover the true arrivals in the presence of a large amount of false picks under low SNR conditions.

In this paper, we continue the study of applying {RANSAC} for picked arrival times and improve our method described in (\citealt{Zhu2016}) for realistic seismic monitoring scenarios. To reduce the false picks in time picking results and thus improve phase association and event location estimation, we propose a {RANSAC}-based arrival time clustering (\acronym) method as a pre-processing step that groups true picked times into different events and identifies false picks. 
A similar approach has recently been used to solve the phase association problem in Chile (\citealt{Woollam2020}). The primary difference between our \acronym method and other classic methods such as GA, GLASS3 (e.g., \citealt{Lebras1994}; \citealt{Yeck2019}) is that we focus primarily on local scale, while the rest focus primarily on global scale or a combination of global, region and local scales.
To demonstrate the accuracy of \acronym\ for 1-D receiver arrays, synthetic simulations are conducted in homogeneous, layered and non-layered isotropic media.
The proposed scheme is also validated through a natural earthquake dataset collected on a 5200-element 2-D surface network in Long Beach, CA.
All cases show that the \acronym\ framework is accurate and robust under low SNR conditions and applicable to a variety of different monitoring setups.

\section{Motivation}

\label{sec:motivation}


To eliminate false picks generated by a time picking algorithm due to background noise, a classifier for true event picks is necessary.
Such a classifier needs to learn the pattern of a seismic event from all arrival-time picks and apply a rule to cluster the picks into two groups: true event and false picks.
It also needs to be robust enough to accommodate the variety of patterns shown by different events.
Since the true first-arrival times of any isolated seismic event result in a predictable moveout curve on a monitoring receiver array, a parametric model for valid moveouts can be used to build a classifier for true picks of an actual seismic event.

Moveout curves have been studied extensively in seismology by \cite{Dix1955} and \cite*{Dellinger1993} (See Appendix A).
For homogeneous media, it is simply a hyperbola.
\cite{Dix1955} proved that moveout curves can also be modeled as hyperbolas for isotropic layered media when the receivers have small offsets relative to an event epicenter.
He also gave explicit parametric equations for such curves (as a rotated hyperbola) when a tilting layer is present.
\cite{Dellinger1993} showed that for transverse isotropic (TI) media, an elliptic parametric model can be used to approximate the expected moveout curves.
Since horizontal variation in velocity is relatively small in microseismic monitoring, a hyperbola can be used to approximate the arrival time moveout curve for an event in non-layered media as well.
To sum up, for a surface monitoring receiver array in microseismic monitoring, a quadratic parametric model exists for a moveout curve observed on a receiver array from a valid seismic event.
Thus, the problem of finding the true picks for a seismic event can be solved by fitting a parametric model using picked arrival times.
For simplicity, we only consider isotropic media with short offsets in this study, which corresponds to a hyperbola.
Ellipses share the same quadratic model as hyperbolas, but with different parameter requirements.

However, due to poor SNR on surface monitoring arrays, there are many false picks that are far from true event moveout curves which we refer to as outliers.
Curve fitting like least squares uses as many data points as possible to minimize the amount of misfit error.
Although it is the optimal solution under a Gaussian random noise assumption, it fails dramatically in the presence of outliers which are unlikely to be Gaussian distributed.
We, instead, adopt a random sampling scheme that repeatedly uses a minimum number of data points to fit a curve, 
and selects the best curve (hypothesized model) as the one with the most data points close to it, i.e., with the maximum number of inliers.

This random sampling scheme was first proposed by \cite{FisBol1981} as random sample consensus {RANSAC} and then improved by many others (\citealt{Stewart1995}; \citealt{Torr2000}; \citealt{Chum2002}; \citealt{TorMur2005}; \citealt{Chum2005}).
It has also been extended to fit multiple models simultaneously (\citealt{Wang2004}; \citealt{Toldo2008}).
Based on the fitted hyperbolic model, the picked arrival times are, in fact, clustered into event groups and non-event groups.
Such clustering not only separates picked arrival times into different phases, e.g., P-wave and S-wave phases, but also improves the accuracy of subsequent source-location estimates by eliminating false picks due to noise.

\section{Method}

\label{sec:method}

\subsection{{RANSAC} overview}

\label{subsec:rasnac}

Despite many variations and adaptations of RANSAC (\citealt*{Choi2009}), there are essentially two steps per iteration (hypothesize-and-test) which will be repeated to yield the best fit to the data:
\begin{itemize}
  \item \emph{Hypothesize}: A \textbf{minimal} sample subset (MinSet, denoted as $\Omega_\mathrm{M}^k$) is randomly selected from the dataset and the \textbf{unique} model parameters ($\mathbf{p}^k$) are computed for $\Omega_\mathrm{M}^k$.
  \item \emph{Test}: Elements in the dataset ($\Omega_\mathrm{D}$) are evaluated to determine which ones can be labeled as \emph{inliers}, i.e., consistent with the hypothesized model in the sense that the distance from the model's moveout curve is less than some prescribed value ($\delta$).
    The set of all such inliers is called a consensus set (ConSet, denoted as $\Omega_\mathrm{C}^k$).
\end{itemize}
Note that $\Omega_\mathrm{M}^k \subset \Omega_\mathrm{C}^k \subset \Omega_\mathrm{D}$.
A set $\Omega_\mathrm{M}^k$ consists of only the minimal number of samples required to uniquely determine a model, e.g., two samples for a line and three for a circle.
The more elements in $\Omega_\mathrm{C}^k$, the better the model we have obtained for the $k^\text{th}$ hypothesis.

Notice that each {RANSAC} iteration requires very little computation and there exists a unique solution for each chosen MinSet.
In this way, we can afford using a large number of iterations consistently to perfect a hypothesized model.
\cite{FisBol1981} give a statistical analysis of the required number of iterations for the {RANSAC} process with an inlier ratio of $u$ (ratio between number of inliers and total number of data points).
The number of iterations $\hat{N}$ to guarantee that at least one $\Omega_\mathrm{M}$ will only contain true picks with probability $p$ is
\begin{equation}
  \hat{N} = \frac{\log(1-p)}{\log(1-u^m)},
  \label{eq:ransac_iteration}
\end{equation}
where $m$ is the size of a MinSet, which is 5 for a hyperbola and 9 for a hyperboloid.
For example, when 50\% of all picks are close to a hyperbola ($u = 0.5, m = 5$), we can guarantee a 99\% chance of finding an outlier-free $\Omega_\mathrm{M}$ ($p = 0.99$), if we run $\hat{N} = 145$ iterations.
For a 2-D surface array, $m = 9$, and then $\hat{N} = 2356$.
Although $\hat{N}$ may be large, the core operations of deriving the hyperbola parameters and testing its validity are extremely fast so thousands of trials are reasonable.

\begin{figure}
  \centering
  \includegraphics[width=0.9\linewidth]{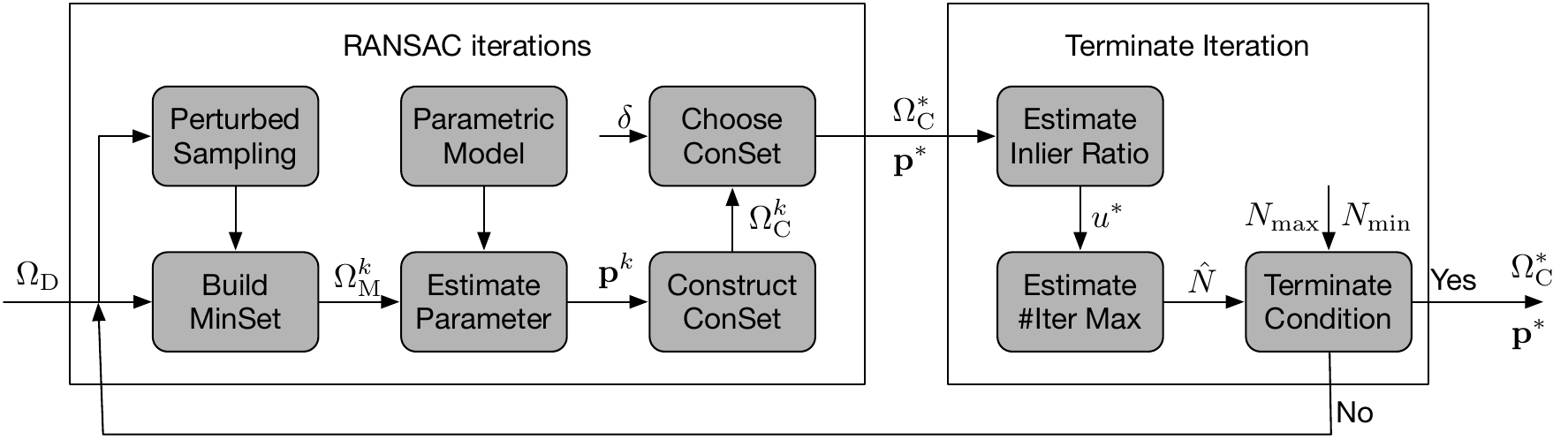}
  \caption{Flow chart of RANSAC process within each iteration. After many iterations the case with the biggest ConSet is declared the best parameter vector $\mathbf{p}^*$. }
  \label{fig:ransac_flow}
\end{figure}

The {RANSAC} process is summarized as the flow chart in Figure~\ref{fig:ransac_flow}.
After the $k^\text{th}$ iteration, the current best ConSet $\Omega_\mathrm{C}^*$ is updated with the $k^\text{th}$ ConSet if $\Omega_\mathrm{C}^k$ has more inliers.
Then $\Omega_\mathrm{C}^*$ is used to estimate the current best inlier ratio $u^*$.
Based on equation \eqref{eq:ransac_iteration}, the number of iterations required $\hat{N}$ can be updated (\citealt{TorMur2005}).
The current $\hat{N}$ is also compared against preset minimum and maximum values $N_\text{max}$ and $N_\text{min}$.
Once the termination condition is satisfied, the best ConSet $\Omega_\mathrm{C}^*$ and model parameter $\mathbf{p}^*$ will be returned; otherwise, the iteration loop will continue.

\subsection{Parameter estimation}

\label{subsec:param_est}

The proposed method uses a quadratic model to estimate the parameters of a hyperbolic curve, which takes the following form:
\begin{equation}
  \mathcal{P}(x, y; \mathbf{p}) =
  \renewcommand\arraystretch{0.5}\begin{bmatrix}
    x & y
  \end{bmatrix}
  \begin{bmatrix}
    a & b/2\\ b/2 & c
  \end{bmatrix}
  \begin{bmatrix}
    x\\[1ex] y
  \end{bmatrix} +
  \begin{bmatrix}
    d & e
  \end{bmatrix}
  \begin{bmatrix}
    x\\[1ex] y
  \end{bmatrix}+ f = 0
  \label{eq:quad_eq_mtx}
\end{equation}
or, equivalently,
\begin{equation}
  \mathcal{P}(x, y; \mathbf{p}) = a x^2 + b xy + c y^2 + d x + e y + f = 0,
  \label{eq:quad_eq}
\end{equation}
where $\mathbf{p}$ has six real elements $\boldsymbol{\theta} = (a, \cdots, f)$, but there are actually only five free parameters, since one of the nonzero elements can be always normalized to 1.

We need to check two determinants in the QC process to guarantee that the quadratic form is a non-degenerate hyperbola.
First, when the determinant
\begin{equation}
  \Delta_1 = 4\left|\renewcommand\arraystretch{0.5}\begin{array}{ccc}
    a & b/2 & d/2\\
    b/2 & c & e/2\\
    d/2 & e/2 & f
  \end{array} \right|
  \label{eq:discriminant1}
\end{equation}
is nonzero ($\Delta_1 \neq 0$), equation \eqref{eq:quad_eq} defines a non-degenerate conic section.
To verify it is hyperbola, we must then check a second determinant
\begin{equation}
  \Delta_2 = 4\left|\renewcommand\arraystretch{0.5}\begin{array}{cc}
    a & b/2\\ b/2 & c
  \end{array} \right| = b^2 - 4ac.
  \label{eq:discriminant2}
\end{equation}
When $\Delta_2 > 0$, equation \eqref{eq:quad_eq} defines a hyperbola.

Given a set of $n$ arrival time picks, $(x_i, y_i)$ for $i=1,\ldots,n$,  we form the $n\!\times\!6$ data matrix $\mathbf{D}_n$ and a $6\!\times\!1$ coefficient vector $\mathbf{p}$, such that $\mathbf{D}_n \mathbf{p}$ is the model $\mathcal{P}(x, y; \mathbf{p})$ in \eqref{eq:quad_eq} evaluated at the time picks.
With measurement error, there will be a nonzero residual $\mathbf{r}$ as follows:
\begin{equation}
  \left[\renewcommand\arraystretch{0.95}\begin{array}{*{6}{c}}
      x_1^2 & x_1^{\phantom{.}} y_1^{\phantom{.}} & y_1^2 & x_1^{\phantom{.}} & y_1^{\phantom{.}} & 1\\
      \vdots & \vdots & \vdots & \vdots & \vdots & \vdots\\
      x_n^2 & x_n^{\phantom{.}} y_n^{\phantom{.}} & y_n^2 & x_n^{\phantom{.}} & y_n^{\phantom{.}} & 1\\
    \end{array} \right] \left[ \renewcommand\arraystretch{0.5}\begin{array}{c}a\\b\\c\\d\\e\\f\end{array} \right] = \left[ \renewcommand\arraystretch{0.95}\begin{array}{c}r_1\\\vdots\\r_n^{\phantom{.}}\end{array} \right]  \quad\Leftrightarrow\quad  \mathbf{D}_n \mathbf{p} = \mathbf{r}.
    \label{eq:design_mtx0}
\end{equation}
We use only five picks to uniquely determine $\mathbf{p}$ as discussed previously.
If all picks in $\Omega_\mathrm{M}$ are true picks, the residual term $\mathbf{r}$ is usually negligible.
Then we solve the linear system $\bm{\mathbf{D}}_5\mathbf{p} = \bm{0}$, which is effectively finding the null space of $\mathbf{D}_5$.
From the singular value decomposition (SVD) of $\mathbf{D}_5$, it is easy to see that the last right singular vector $\mathbf{v}_6\in \text{null}(\mathbf{D}_5)$.
Comparing with the pseudo-inverse method used in (\citealt{Zhu2016}), the solution $\mathbf{v}_6$ adopted here is not guaranteed to have the minimal $L_2$ norm.
However, the SVD approach avoids numerical stability problems when the matrix $\mathbf{D}_n\mathbf{D}_n^T$ is ill-conditioned.
To process the large number of candidate MinSets $\Omega_\mathrm{M}$ efficiently, we do a quality control (QC) of $\mathbf{p}$ by checking determinants, $\Delta_1\neq 0$ and $\Delta_2 > 0$, before proceeding to the more computationally demanding test step that computes the distance between the entire set of picks $\Omega_\mathrm{D}$ and the hypothesized RANSAC model to obtain a ConSet $\Omega_\mathrm{c}$.

\subsection{Add perturbation to MinSets}

\label{subsec:perturbate}

\begin{figure}
  \centering
  \begin{subfigure}[b]{0.48\textwidth}
    \centering
    \includegraphics[width=\textwidth]{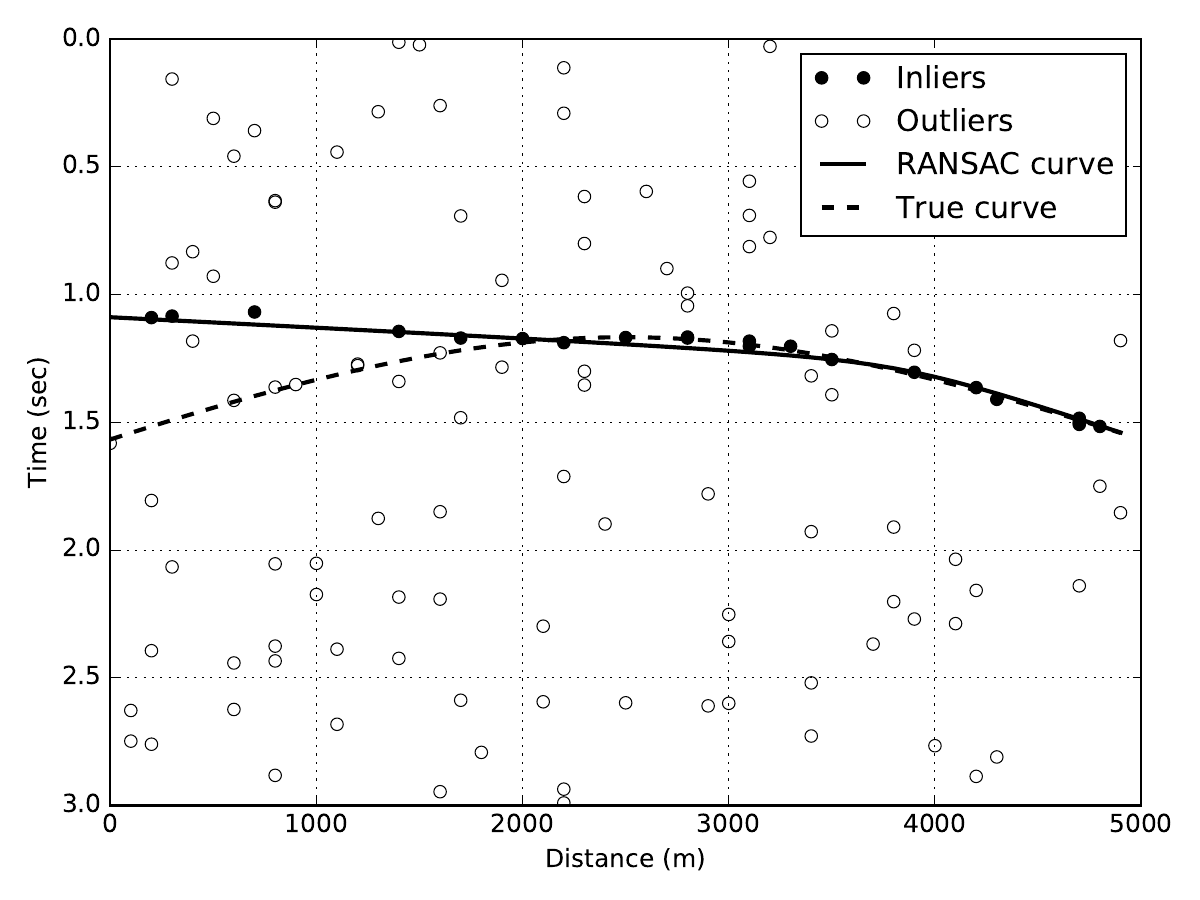}
    \caption[]%
    {{\small }}
    \label{fig:bad_fit}
  \end{subfigure}
  \hfill
  \begin{subfigure}[b]{0.48\textwidth}
    \centering
    \includegraphics[width=\textwidth]{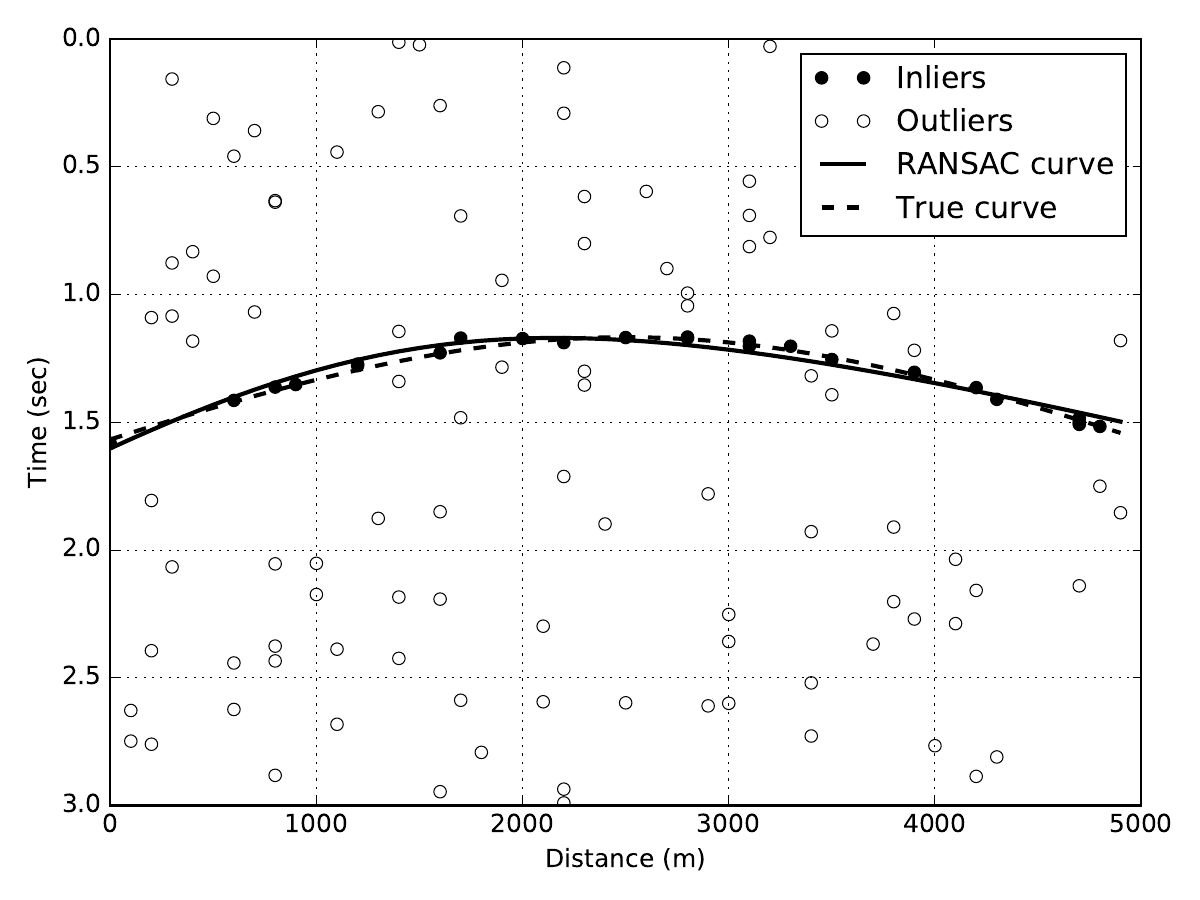}
    \caption[]%
    {{\small }}
    \label{fig:good_fit}
  \end{subfigure}
  \caption[]
  {\small RANSAC classification of noisy data: (a) suboptimal fitting results from  direct null space method, (b) optimal fitting results from null space method with perturbations.}
  \label{fig:ransac_noise}
\end{figure}

With the presence of measurement noise, the null-space method has a tendency to fit the wrong type of curve, namely a parabola or an ellipse, which is then eliminated by the parameter QC step.
In noisy cases when the percentage of inliers is low, this problem may cause RANSAC to select a suboptimal parameter vector whose curve passes through some outliers as well as true inliers, as shown in Figure~\ref{fig:bad_fit}.

To overcome this tendency when fitting quadratic models, a constrained least squares approach (\citealt*{Fit1999}; \citealt{Leary2004}) has been developed to force the fitted curves to be hyperbolas (and ellipses) by solving a generalized eigen system determined by a constraint matrix.
Although this method works for general hyperbolic curve fitting problems, its strategy runs counter to {RANSAC}'s MinSet assumption, i.e., using a random set with the minimum number of points.
Least squares employs as many data points as possible in order to minimize the distance between the data and the optimal model.
However, the more points required in the MinSet, the smaller the possibility that a MinSet will be outlier-free.
This dilemma restricts the ability of the constrained least-squares method 
to find a proper model when the SNR is low.

Fortunately, the RANSAC process has the luxury of running a very large number of fast iterations to find the optimal model.
By randomly perturbing the time picks in $\Omega_\mathrm{M}^k$ to produce a few additional MinSets and running more iterations, we can combat the effects of measurement noise.
For example, after adding a small perturbation to the picked arrival time in Figure~\ref{fig:ransac_noise}a, the correct moveout curve is selected in Figure~\ref{fig:ransac_noise}b.
Although we perturbed the picked arrival times to get the model coefficients, the final output of inliers and the location estimation are conducted on the original ``unperturbed'' data.

\subsection{Processing pipeline}

\begin{figure}
  \centering
  \includegraphics[height=0.6\textheight]{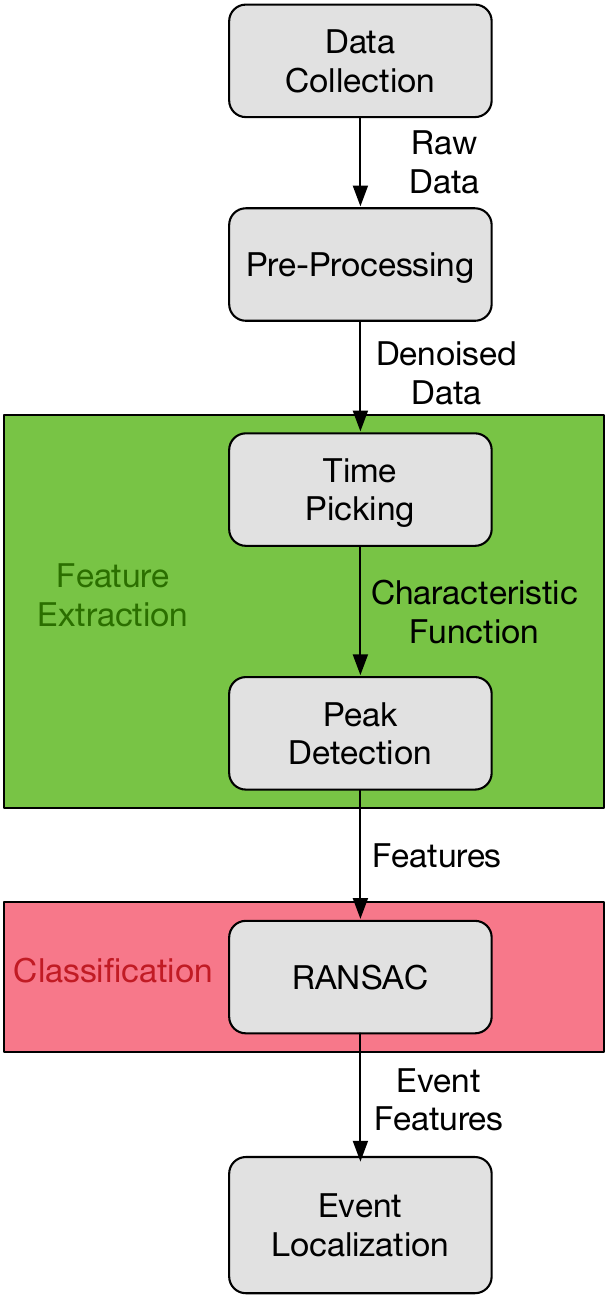}
  \caption{Processing pipeline of proposed \acronym\ method: peak detection can be customized to adapt to the RANSAC framework; classification based on RANSAC eliminates false picks when estimating the moveout curve for an event.}
  \label{fig:pipeline}
\end{figure}

The overall processing pipeline of \acronym\ is summarized in Figure~\ref{fig:pipeline}.
Arrival times are picked on pre-processed data to extract event features out of seismic traces as time pick pairs $(\mathbf{x}, t)$.
In this study, we use the widely adopted short-term over long-term average ratio (STA/LTA) method to generate a characteristic function for each seismic trace.
However, any valid time picking method, such as those included in (\citealt{AkramEaton2016}), can replace STA/LTA depending on the specific SNR condition.
Peak detection is conducted on characteristic functions to generate $(\mathbf{x}, t)$ pairs which are then clustered by {RANSAC} to select true picks that correspond to a valid event moveout. These clustered picks can be fed into other location estimation programs such as double difference (\citealt{Wald2000}; \citealt{Zhang2003}).
When no prior knowledge of the velocity model is available, we provide a moveout curve fitting based event location estimator assuming a homogeneous medium.

Notice that this is a highly flexible framework in which multiple methods can be used for each block to optimize the performance for different datasets.
Figure~\ref{fig:pipeline} provides a generic approach to demonstrate the accuracy and robustness of \acronym; however, it can be customized to specific needs and easily incorporated into any time picking based processing workflow.

\section{Pre-process seismic data for {RANSAC} classification}

\label{sec:preprocessing}

The input data can be pre-processed to facilitate peak detection and help {RANSAC} better fit the moveout curve.
The strategy is to encourage more time peaks by including as many weak events as possible while not introducing too many false picks.
Here we give an example pre-processing method that takes advantage of {RANSAC}'s ability to eliminate outliers while including more weak picks that might be related to a true event.

\subsection{Guided peak detector}

\label{subsec:peak_detector}

\begin{figure}
  \centering
  \includegraphics[width=\linewidth]{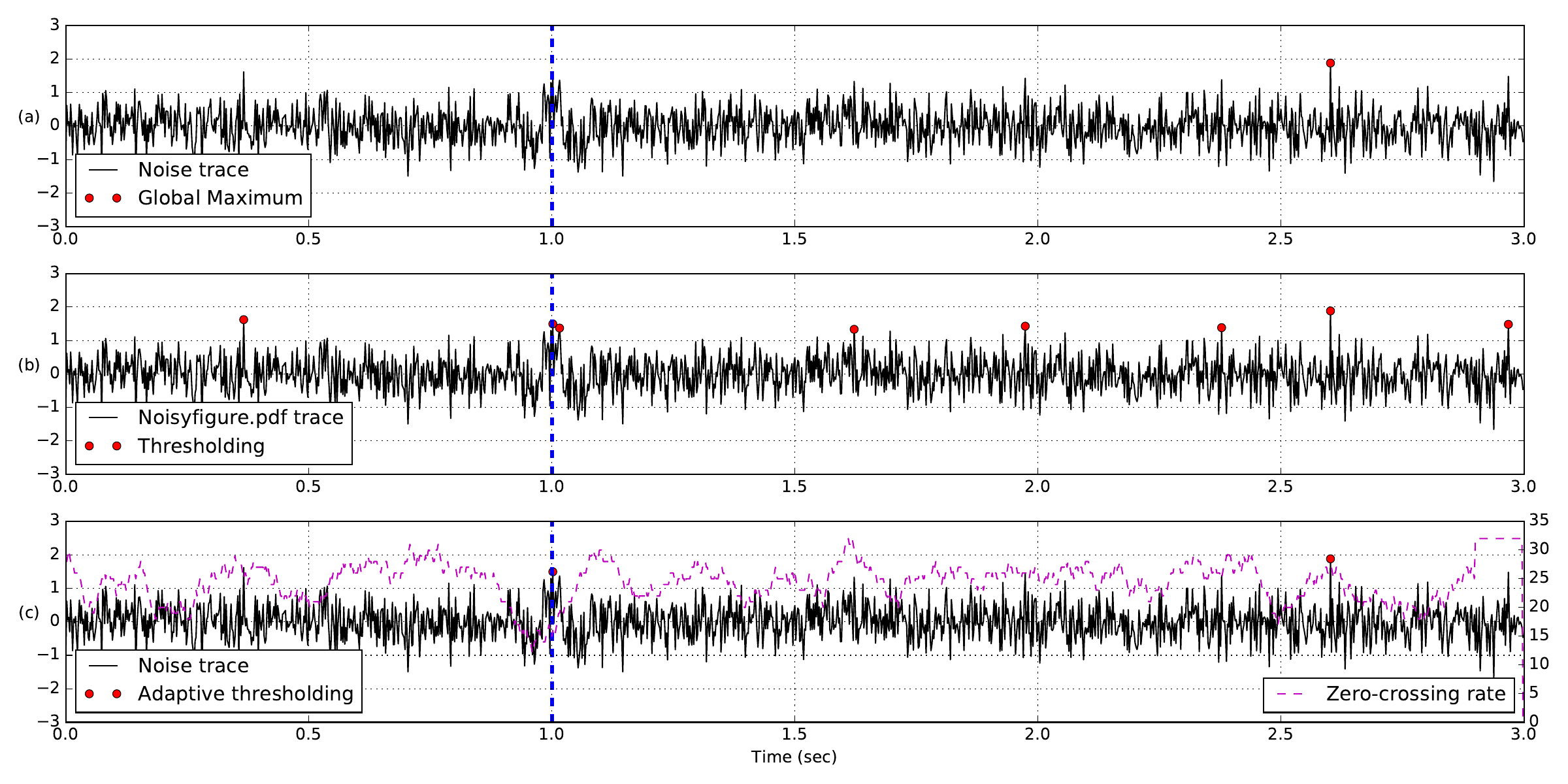}
  \caption{Detection methods for single event case on noise trace with PSNR = 6\,dB (true peak at $t=1.0$\,sec): (a) global maximum; (b) thresholding at 70\% of global maximum; and (c) adaptive thresholding at 95\% of global maximum weighted by the local zero-crossing rate. Adaptive thresholding picks true arrival time cluster with only one false pick.
  }
  \label{fig:peak_detection}
\end{figure}

A straightforward approach to peak detection in characteristic functions is to find the global maximum on each trace.
However, such peak locations can easily be affected by background noise as shown in Figure~\ref{fig:peak_detection}a, and smaller events are overlooked when multiple events are present.
Likewise, locating peaks by local maxima is adversely affected by background noise since a noise signal tends to have a large number of local peaks.

One way to include more time picks is to use thresholding at a fraction of the maximum value.
Shown in Figure~\ref{fig:peak_detection}b, setting the threshold to 70\% of the maximum value yields more picks.
This method succeeds in finding more arrival times (both true and false), but puts a heavy burden on the following classification block if too many of these peaks are false ones.
A better way to include more time picks is to make a rough estimate of where the real signal lies based on a signal attribute such as the local zero-crossing rate defined below:
\begin{equation}
  r_{\text{zc}}(\tau) = \frac{1}{T}\sum_{t=\tau-T/2}^{\tau+T/2}\mathbf{1}(\mathbf{s}_t\, \mathbf{s}_{t-1} < 0),
  \label{eq:zcr}
\end{equation}
where $\mathbf{1}(\mathbf{s}_t\, \mathbf{s}_{t-1} < 0)$ counts sign changes in	$\mathbf{s}_t = \text{sgn}(s(t))$, and $T$ is the interval over which $r_{\text{zc}}(\tau)$ is computed.
The local zero-crossing rate should be low when the signal is present.
Using 95\% of the global maximum threshold on the characteristic function weighted by $\sqrt{r_{\text{zc}}(\tau)}$, 
Figure~\ref{fig:peak_detection}c shows that this guided peak detector successfully picks only the real arrival time peak and the global maximum (due to severe background noise).

\subsection{Merging close picks}

\label{subsec:peak_merger}

With background random noise, there may be multiple peaks clustered around a true event pick.
This not only leads to more computational cost in later location algorithms but also introduces uncertainty into event locations.
Such a problem can be solved by merging close picks into one pick.
A common practice in manual picking is to use the starting point of a pick cluster as the event arrival time.
This is reasonable as the peak detector usually picks both the arrival signal (first break) and its coda wave (points that follow which form a cluster of time picks).
However, it not only requires more computation to search for closely located peaks but also can be misled by a false pick that slightly leads the true picks.
It can soon become tricky to set the correct parameter for how close the picks need to be to each other for a merge.

We consider using Gaussian smoothing which is widely used for edge detection in image processing (\citealt{Basu2002}).
Gaussian smoothing helps in reducing details (adjacent small peaks) within the characteristic function and attenuating insignificant local peaks due to noise.
It convolves the response function with a Gaussian function defined below:
\begin{equation}
  G(x) = \frac{1}{\sqrt{2\pi\sigma^2}}e^{-\frac{x^2}{2\sigma^2}},
\end{equation}
where $\sigma$ is the standard deviation which can be set as the dominant duration of a wavelet (0.1\,sec in this case).
Although Gaussian smoothing cannot eliminate all the close false picks, it can help mitigate such errors.
Note that Gaussian smoothing is only helpful for traditional phase picking methods, e.g. STA/LTA, which are sensitive to local peaks.
It should be omitted when using phase picking method based on full waveforms, such as cross-correlation (\citealt{song2010improved}) or neural networks (\citealt{zhu2019deep}).

\section{Seismic examples}

\label{sec:examples}

In this section, we explore the performance of the proposed \acronym\ method in a more realistic scenario of seismic processing.
In the first example, a Ricker wavelet is manually delayed with moveout from a homogeneous medium assumption to demonstrate the essence of the proposed method.
The second example uses a recorded seismic trace consisting of P-wave and S-wave phases which are then manually delayed according to a layered velocity model to simulate data from an array.
This is a typical scenario in microseismic surface monitoring and we demonstrate that \acronym\ is able to extract both P-wave and S-wave phases and group them into event clusters.
In the third example, we explore the problem when the layered model assumption is violated by using the Marmousi2 velocity model to generate the testing data with a finite-difference time-domain (FDTD) simulation.
In the final example, we demonstrate that \acronym\ can be easily extended to the case of a 2-D surface monitoring array by validating it on a 5200-element 2-D dense array deployed for earthquake monitoring in Long Beach, CA.

\subsection{Ricker wavelet in homogeneous media}
\label{subsec:ricker_homo}

\begin{figure}
  \centering
  \includegraphics[width=\textwidth]{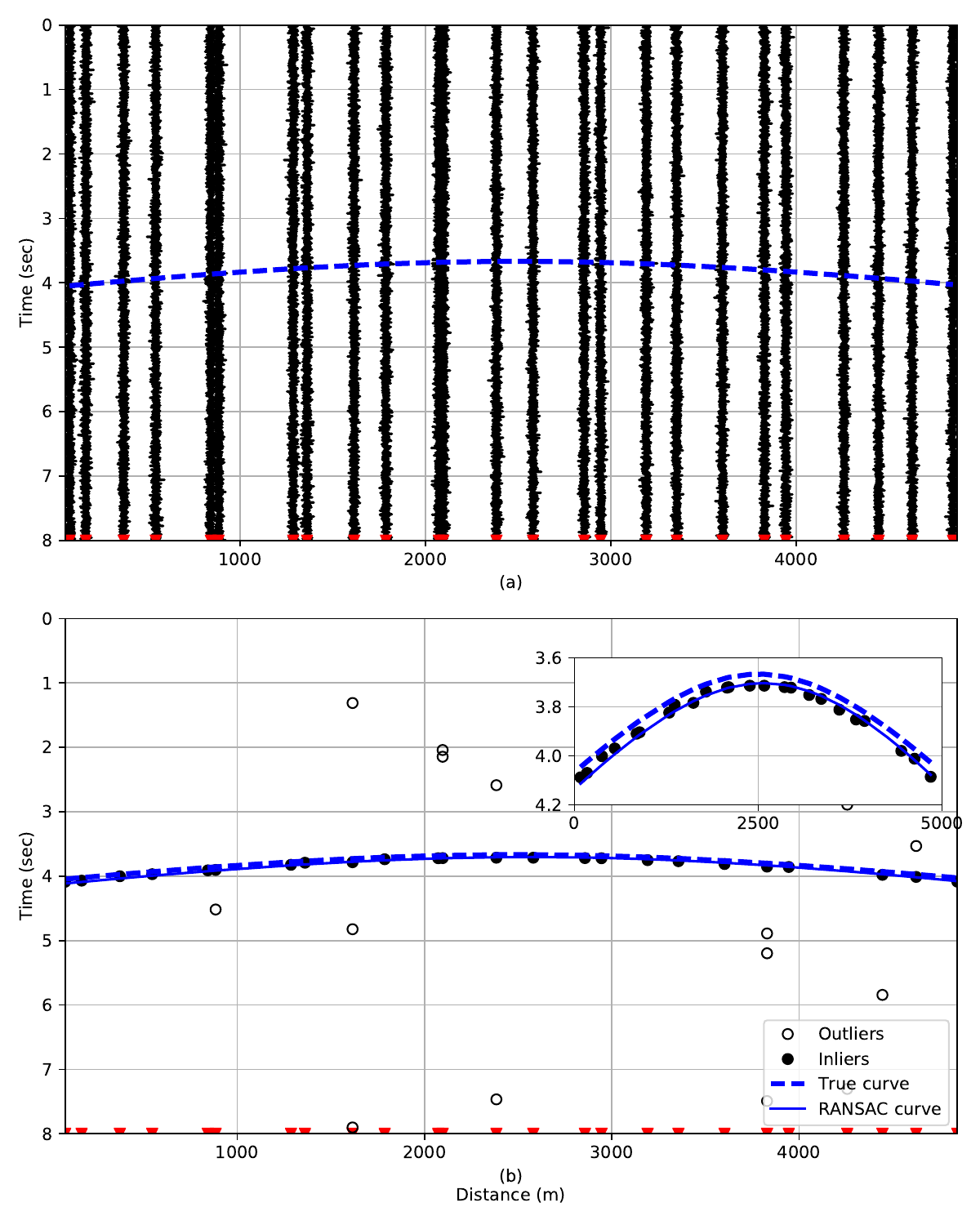}
  \caption{Simple example of time picking: (a) raw traces of a 25-element nonuniform linear array with an aperture of 5\,km. Source waveform is a Ricker wavelet of 10\,Hz central frequency and the PSNR = 6\,dB. Blue dashed curve indicates the true moveout. (b) Arrival times picked on each trace clustered into inlier (solid dots) and outlier (empty dots) groups. Insert in the top right shows a zoomed-in view along the vertical (time) axis.}
  \label{fig:simple_example}
\end{figure}

For Figure~\ref{fig:simple_example}a, a 25-element linear array with nominal spacing of 200\,m is deployed on the surface (i.e., 5\,km aperture) to monitor a deep event 2\,km below the array center.
The receiver locations are perturbed by additive white Gaussian noise (AWGN) with $\sigma\!=\!50$\,m to simulate receiver offsets in the field away from uniformly spaced locations due to unavoidable physical restrictions in the field.
Note that such perturbations effectively create a nonuniform linear array.
The raw data section is shown in Figure~\ref{fig:simple_example}a with the true moveout curve marked with a blue dashed line.

The source wavelet employed here is a Ricker wavelet, and the medium is assumed to be homogeneous with a velocity of 3\,km/s.
AWGN with peak signal-to-noise ratio (PSNR) of 6\,dB is added to simulate random background noise.
Because PSNR is not affected by the trace length,
it is used to measure the noise level throughout the paper.
Its definition is as follows:
\begin{equation}
  \text{PSNR} = 20\log_{10}\frac{\max(|s_i(t)|)}{\sigma},
\end{equation}
where $s_i(t)$ is the signal at the $i^\text{th}$ receiver, and $\sigma$ is the standard deviation of the AWGN.

After applying the STA/LTA method on each trace, we use the zero-crossing guided peak detector and peak merging to find the candidate arrival time picks.
These picks are passed to a classification block to be grouped into event (inlier) and non-event (outlier) clusters.
When zoomed in around the moveout curve as shown in Figure~\ref{fig:simple_example}b, a small deviation between the fitted curve and the true moveout curve is observed; however, all picks close to the true moveout curve are successfully clustered into the event/inlier group.

Once clustering and correction are complete in the previous steps, the improved picked arrival times in this example can be used to locate events.
There are many existing event location methods that use picked arrival times, such as Geiger's method (\citealt{Geiger1912}) and the double-difference method (\citealt{Wald2000}; \citealt{Zhang2003}).
The corrected arrival times can be used by these methods with known velocity models to improve the location estimation.

When the velocity model is unknown, we can assume a homogeneous medium in order to compute predicted arrival times from possible source locations.
Based on the inliers given by \acronym, we can minimize a nonlinear objective function that measures the sum of squared errors between the \acronym\ classified inlier picks and the predicted arrival times
\begin{equation}
  \epsilon = \sum_{i=1}^n (t_i - t_i^p(\mathbf{x}, T_0, v))^2
  \label{eq:LocationViaTravelTimeMatching}
\end{equation}
where $t_i$ is the \acronym\ pick at the $i^\text{th}$ receiver and $t_i^p$ is the predicted arrival time which is a hyperbola that depends on the event location $\mathbf{x}$, origin time $T_0$, and homogeneous velocity $v$.
The minimizer of equation \eqref{eq:LocationViaTravelTimeMatching} gives the event location and event origin time, and the medium velocity simultaneously.

\begin{figure}
  \centering
  \begin{minipage}{0.85\linewidth}
    \centering
    \includegraphics[width=\linewidth]{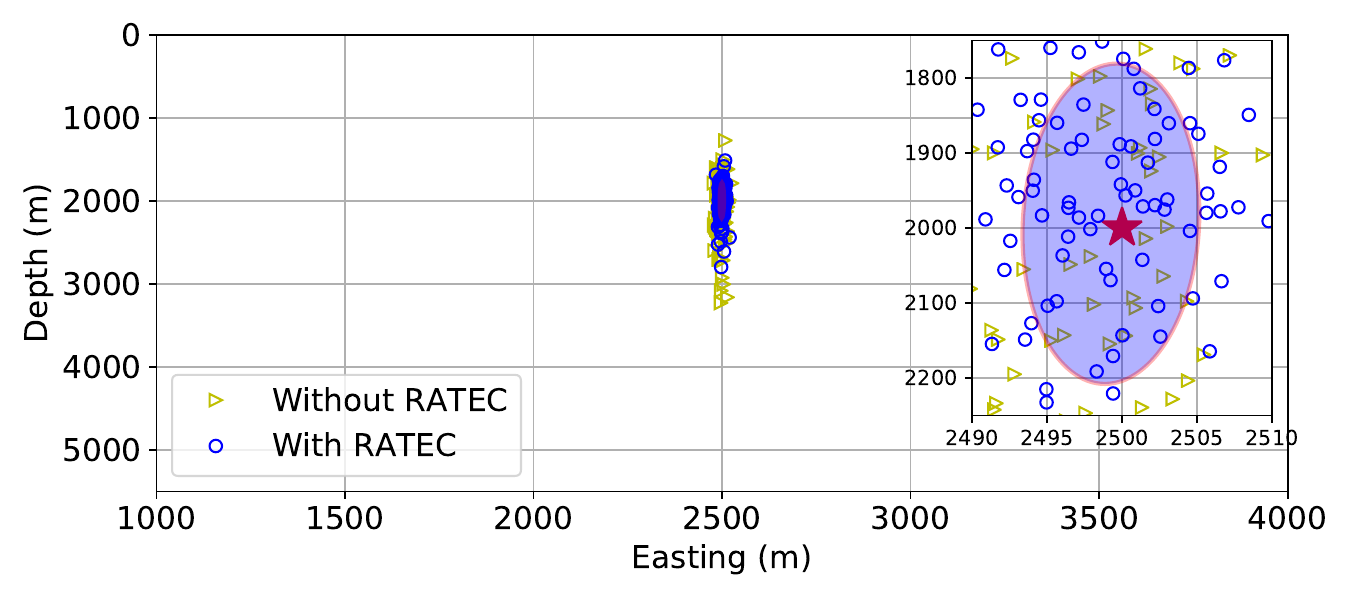}
  \end{minipage}\\
  (a)\\
  \begin{minipage}{0.85\linewidth}
    \centering
    \includegraphics[width=\linewidth]{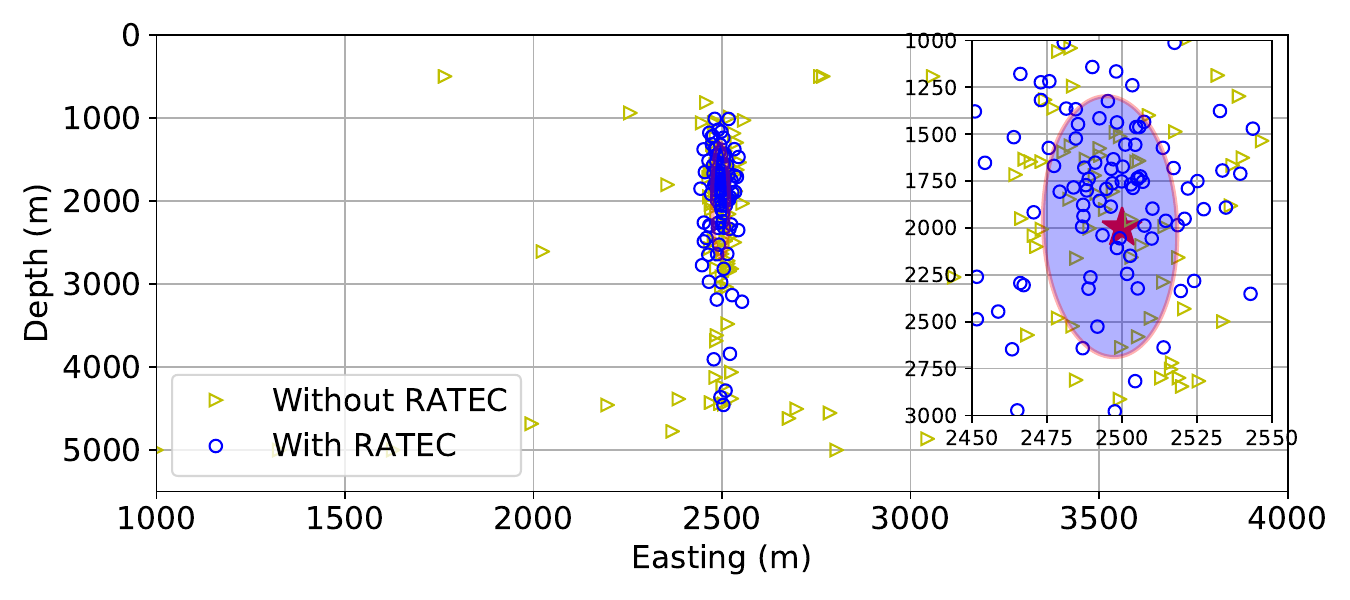}
  \end{minipage}\\
  (b)\\
  \begin{minipage}{0.85\linewidth}
    \centering
    \includegraphics[width=\linewidth]{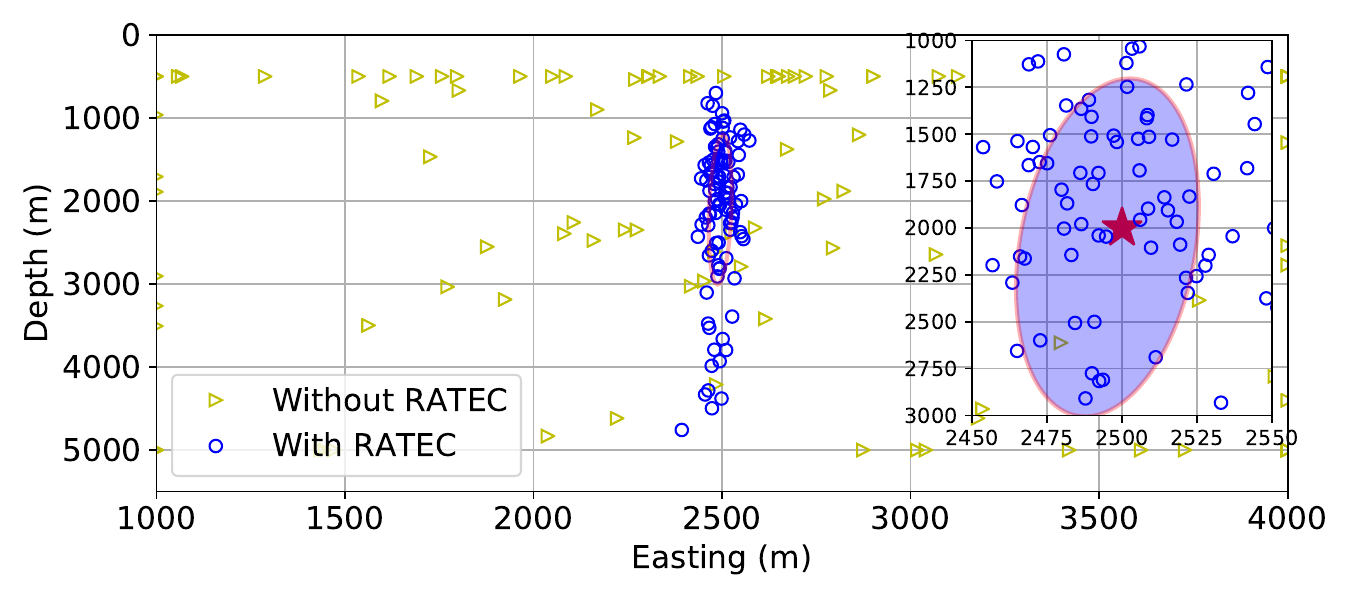}
  \end{minipage}\\
  (c)
  \caption{Location results with and without the \acronym\ scheme using 100 Monte Carlo experiments under three different background PSNR noise levels: (a) 20\,dB, (b) 8\,dB, and (c) 6\,dB. Blue ellipses (in the insets) show the one-sigma confidence interval of the event location estimator. Note the changing axis limits for the insets. The red star indicates the minimizer for the noiseless case which overlaps the true event location at $(2500, 2000)$ m.}
  \label{fig:localization}
\end{figure}

To validate the accuracy of this location scheme, 1000 Monte Carlo experiments are conducted to compare the results with and without the \acronym\ scheme under different noise levels.
Only the first 100 data points, which is sufficient to capture the location estimation distribution as a point cloud, are shown in Figure~\ref{fig:localization} to avoid crowding.
For low background noise, PSNR = 20\,dB, both methods obtain the event location accurately around the true event location $(2500, 2000)$\,m as shown in Figure~\ref{fig:localization}a.
The uncertainty in depth is mostly a result of the array geometry which has poor resolution in the direction perpendicular to the linear array.
The location results with \acronym\ have a more compact distribution around the true location.
There are more blue dots than yellow triangles within the one-sigma confidence interval indicated by the blue ellipse.
The location estimates without \acronym\ start to fall apart when the PSNR is around 8\,dB as shown in Figure~\ref{fig:localization}b.
Although most of the yellow triangles are still around the true location region with a larger spread, there are a significant number of location estimates far away from the event region.
On the other hand, the results with \acronym\ show a consistent distribution around the true event region in Figure~\ref{fig:localization}b.
Under severe noise as shown previously in Figure~\ref{fig:simple_example} with 6\,dB PSNR, the location results without \acronym\ become completely unreliable while those with \acronym\ still give very good estimates.
In Figure~\ref{fig:localization}c about 50\% of the blue dots, but only one yellow triangle, lie within the confidence ellipse.

The accuracy of the location estimate with and without \acronym\ measured in root-mean-square error (RMSE) for the complete 1000 Monte Carlo experiments are shown in Table~\ref{tab:localization}.
With the \acronym\ correction, the location estimate in easting is improved significantly.
The error in depth is much larger but is reduced by applying the \acronym\ correction.


\begin{table}[h]
  \centering
  \caption{RMSE of \acronym\ location results from 1000 Monte Carlo experiments.}
  \label{tab:localization}
  \begin{tabular}{c|c|c|c|c|c|c|}
    \cline{1-7}
    \multicolumn{1}{ |c|  }{\multirow{2}{*}{RMSE }} & \multicolumn{3}{c|}{Easting} & \multicolumn{3}{c|}{Depth} \\\cline{2-7}
    \multicolumn{1}{ |c|  }{} &  {20\,dB}& {8\,dB}& {6\,dB}& {20\,dB}& {8\,dB}& {6\,dB}  \\\hline
    \multicolumn{1}{|c|}{Without \acronym\ } & 8.88 & 312.21 & 911.79 & 386.96 & 1273.19 & 1914.61\\
    \cline{1-7}
    \multicolumn{1}{|c|}{With \acronym\ } & 6.09 & 25.49 & 65.73 & 199.43 & 869.78 & 1025.52\\\hline
  \end{tabular}
\end{table}

\subsection{Recorded seismic trace in layered model}

\begin{figure}
  \centering
  \includegraphics[width=\textwidth]{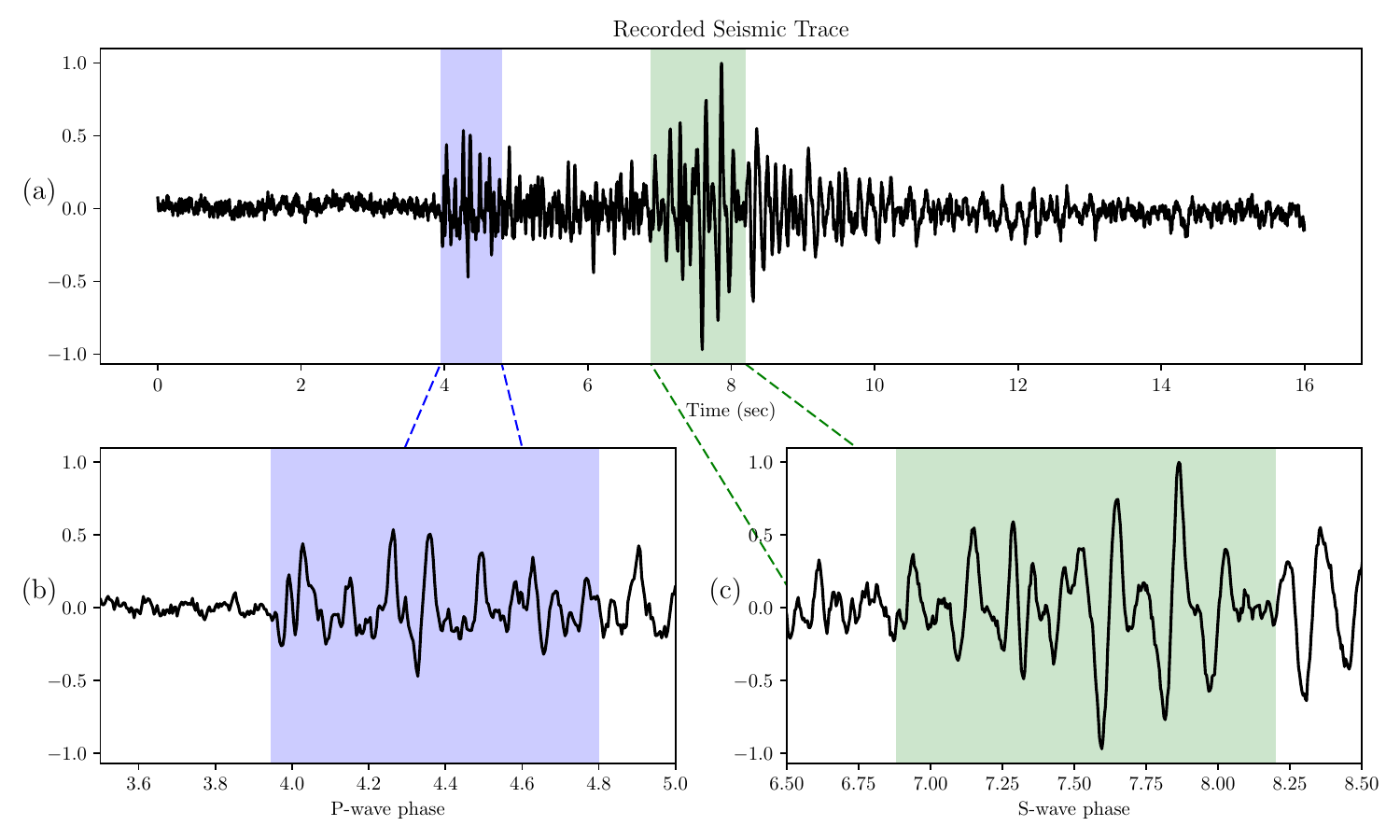}
  \caption{(a) Recorded seismic trace with P-wave (blue) and S-wave (green) phases for layered model simulation, (b) manual pick of P-wave phase, and (c) S-wave phase. The phase picks are the beginning of the marked time windows.}
  \label{fig:real_seis_trace}
\end{figure}

Here, the seismic trace shown in Figure~\ref{fig:real_seis_trace}a is used as a source signal.
After manually picking the P-wave and S-wave phases, shown in Figure~\ref{fig:real_seis_trace}b and \ref{fig:real_seis_trace}c respectively, the P and S phases are delayed separately according to their travel time ($T$) computed from a layered model against horizontal offset ($x$) using the parametric equation \eqref{eq:tt_layered} given by \cite{Dix1955}.
A detailed explanation can be found in Appendix \ref{app:param_model}.
\begin{equation}
  \left\{ \begin{aligned}
      x = \sum_k \frac{ph_kv_k}{\sqrt{1-(v_kp)^2}}\\
      T = \sum_k \frac{h_k}{v_k\sqrt{1-(v_kp)^2}}
  \end{aligned}\right.
  \label{eq:tt_layered}
\end{equation}
where $p$ is the ray parameter that is constant among all layers, $h_k$ and $v_k$ are layer thickness and layer velocity which defines a layered velocity model.
Unlike the hyperbolic approximation discussed before, this equation is mathematically valid even when $x \to \infty$; however, the direct wave may not necessarily be the first arrival wave when $x$ is large.
In addition, for large $x$ cases, the SNR condition may be too bad for a valid location problem.
Thus, all examples here are conducted for small $x$ (less than 5\,km).

\begin{figure}
  \centering
  \includegraphics[width=\textwidth]{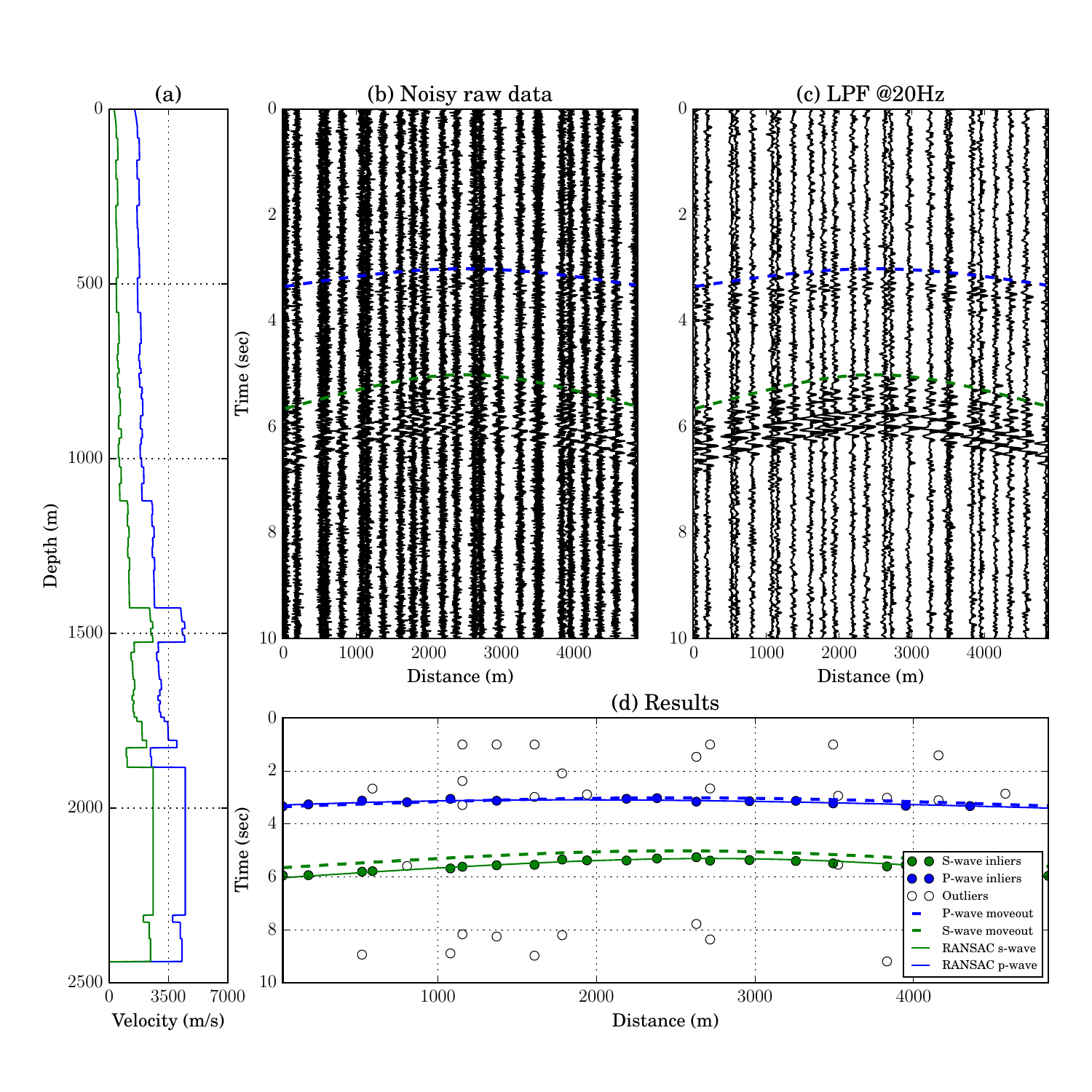}
  \caption{Layered velocity model example using recorded seismic trace with P-wave and S-wave phases: (a) 1-D velocity model from Marmousi2; (b) noisy raw data with PSNR\,=\,10\,dB with respect to the S-wave peak; (c) 20\,Hz low-passed data; (d) fitted moveout curve and classification results comparing to true P-wave and S-wave moveout curves.}
  \label{fig:layered_example}
\end{figure}

The layered velocity model used in this example as shown in Figure~\ref{fig:layered_example}a is taken from Marmousi2 elastic velocity model (\citealt*{Martin2006}).
The top water layer in Marmousi2 is removed and event source is located around 2.5\,km deep.
The same nonuniform surface array as in Section~\ref{subsec:ricker_homo} is used for monitoring underground seismic events occurring at the center of the array.

With noise at 10\,dB PSNR, the P-wave and S-wave arrivals are not obvious in the raw data shown in Figure~\ref{fig:layered_example}b.
With a	spectrogram the dominant frequency of the arrival event is estimated to be 10\,Hz, so a low-pass filter with cutoff frequency at 20\,Hz is used as pre-processing.
Both P-wave and S-wave arrivals are observed in Figure~\ref{fig:layered_example}c after low-pass filtering.
The result of applying the \acronym\ method is shown in Figure~\ref{fig:layered_example}d, where moveout curves were generated by fitting the classified and corrected arrival time picks.
The proposed method is used iteratively in this example to extract all possible event phases: after one moveout curve is detected and identified, its outliers are used as the input for the next iteration to search for more curves until there are not enough time picks to successfully define a moveout curve.
Here, both P-wave (blue) and S-wave (green) phases are identified in this example with most of the true arrival times labeled correctly.

\subsection{Ricker wavelet in non-layered media example}

\begin{figure}
  \centering
  \includegraphics[width=\textwidth]{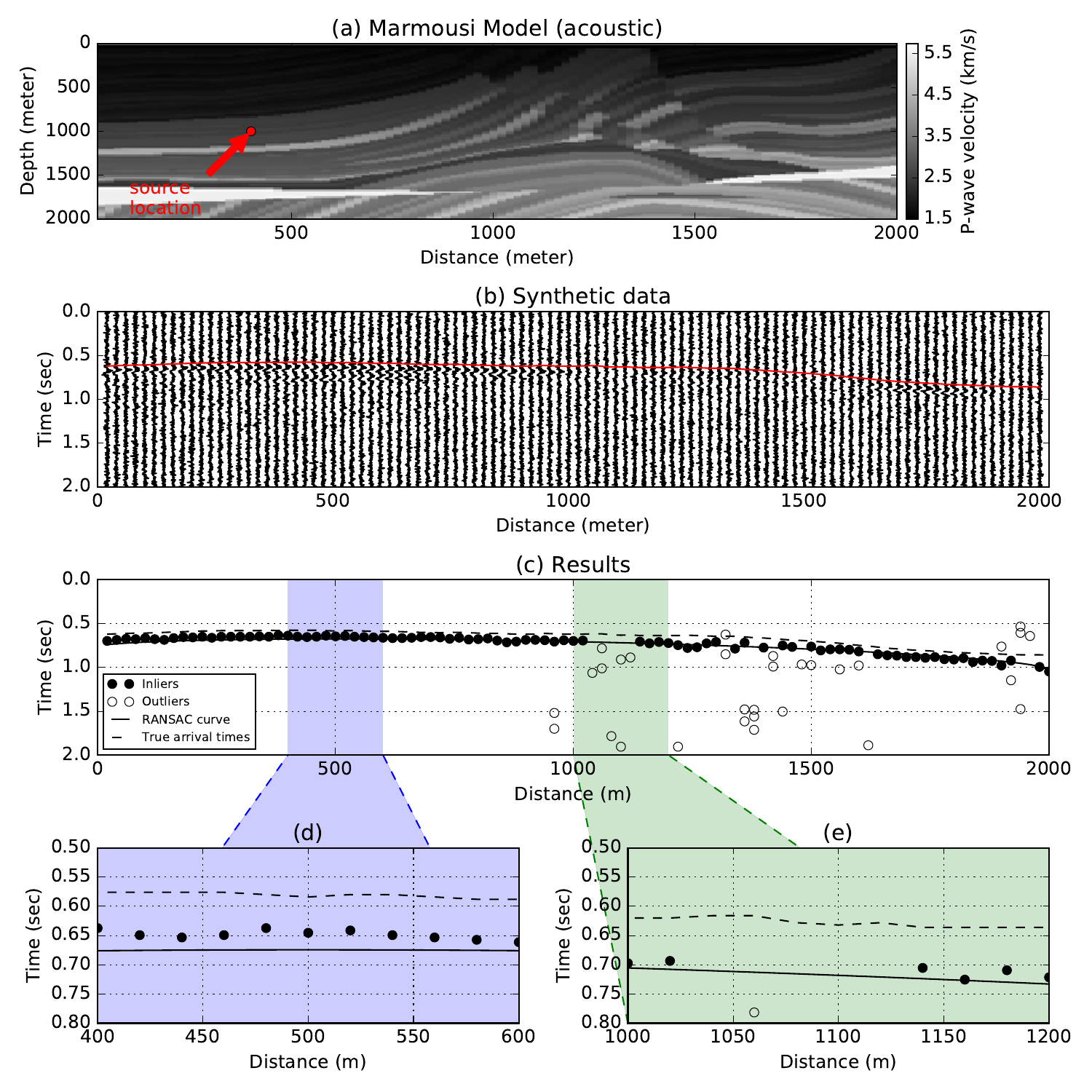}
  \caption{Marmousi model example under 10\,dB PSNR: (a) velocity model (red dot indicates source location), (b) synthetic data (red line indicates true arrival times), (c) \acronym\ results shown as solid line versus true arrivals shown as dashed line, (d) zoomed-in results between 400\,m and 600\,m and (e) zoomed-in results between 1000\,m and 1200\,m.}
  \label{fig:marmousi_example_10dB}
\end{figure}

Although \acronym\ is based on a layered velocity model assumption, it is robust enough to handle non-layered models to some extent.
In Figure~\ref{fig:marmousi_example_10dB}a, the acoustic Marmousi model is used to introduce horizontal variation in the velocity model.
A finite-difference time-domain based numerical simulation is used to generate the surface receiver data shown in Figure~\ref{fig:marmousi_example_10dB}b with 10\,dB PSNR of AWGN.
Since each trace has different peak value, which is common in a real seismic scenario, the PSNR defined here uses the global peak of all the traces.
Receivers in the layered region ($0\sim1000$\,m horizontally) tend to have better SNR than those in the non-layered region ($1000\sim1500$\,m).

After applying the \acronym\ scheme, Figure~\ref{fig:marmousi_example_10dB}c shows the results of curve prediction and arrival time labels.
Even though the true moveout is not exactly a hyperbola, the \acronym\	method is able to label all the true arrival times given a 0.05\,s tolerance range.
Zoomed in around the layered region, good prediction and perfect labeling are observed in Figure~\ref{fig:marmousi_example_10dB}d.
Notice that there now exists larger offsets between picked and true arrival times.
Figure~\ref{fig:marmousi_example_10dB}e shows the results in the non-layered region where the SNR is worse.
Despite the fact that many picks in that region are false picks,	 RANSAC is able to eliminate most of the picks far away from the true moveout curve and label the true time picks correctly.

\subsection{Surface extension on Microearthquake data}

\acronym\ can be easily extended to a surface array by changing the underlying hyperbolic curve model to a hyperboloid surface model.
Similar to equation \eqref{eq:quad_eq_mtx}, a hyperboloid surface can be defined using a 3-D quadratic equation with ten parameters which takes the following general form:
\begin{equation}
  \mathcal{P}(x, y, z; \mathbf{p}) = \left[\renewcommand\arraystretch{0.5}\begin{array}{ccc}x&y&z\end{array} \right]\left[\renewcommand\arraystretch{0.5}\begin{array}{ccc}
    a & b/2 & d/2\\
    b/2 & c & e/2\\
    d/2 & e/2 & f
    \end{array} \right] \left[\renewcommand\arraystretch{0.75}\begin{array}{c}x\\y\\z\end{array}\right] + \left[\renewcommand\arraystretch{0.5}\begin{array}{ccc}
    g & h & i
    \end{array}\right]\left[\renewcommand\arraystretch{0.71}\begin{array}{cc}
    x\\y\\z
  \end{array}\right] + j = 0.
  \label{eq:quad_eq_mtx3d}
\end{equation}
Using the same \acronym\ scheme, we can adapt the framework to hyperboloid surface fitting by finding the parameter vector $\mathbf{p}=(a,b,\ldots,j)$ in a 10-dimensional space.
Although this may seem to be a much larger parameter space, it adds little burden on the search process as RANSAC searches only $\Omega_\mathrm{M}$ rather than complete parameter space.

\begin{figure}[btp!]
  \begin{minipage}{0.5\textwidth}
    \centering
    \includegraphics[width=\columnwidth]{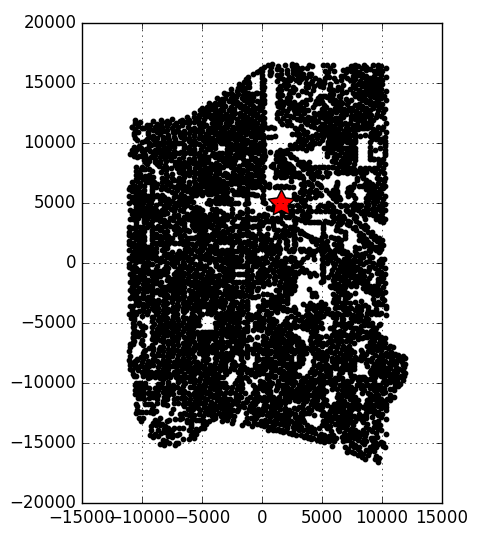}
    \caption{Top view of the sensor array with located event indicated by red star.}
    \label{fig:map_view}
  \end{minipage}%
  \hspace{5mm}
  \begin{minipage}{0.4\textwidth}
    \centering
    \includegraphics[width=\columnwidth]{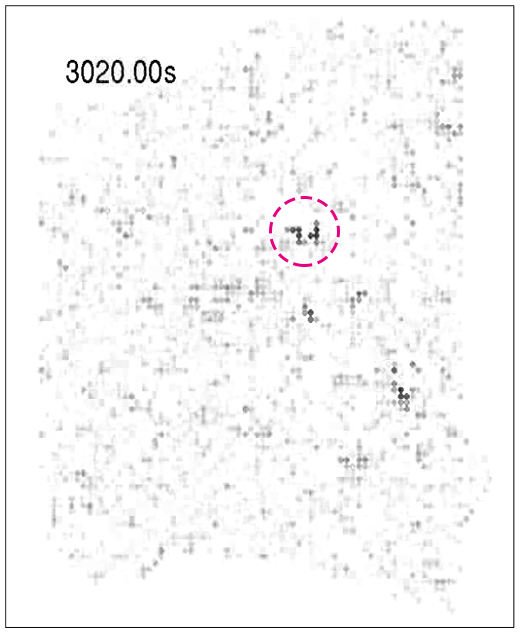}
    \caption{Snapshot of the seismic dataset at time $t=3020.00\,\textrm{s}$;  the visible event lies inside the red circle.}
    \label{fig:sensorSnapshot}
  \end{minipage}
\end{figure}

\begin{figure}
  \centering
  \includegraphics[width=0.55\textwidth]{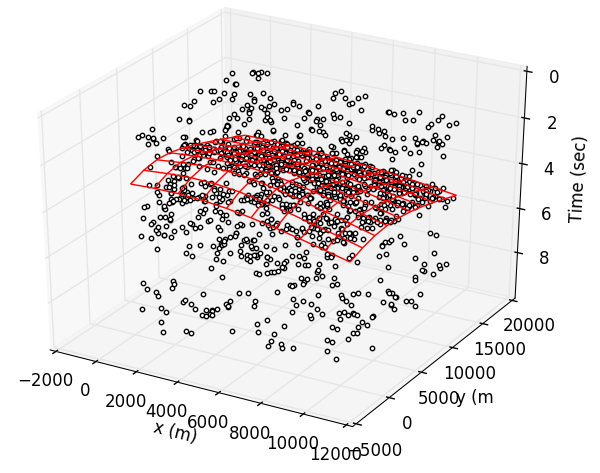}
  \caption{Top view of the picks $(\circ)$ from the 2-D sensor array with fitted surface in red.}
  \label{fig:surface_fit}
\end{figure}

The proposed method was then tested on a data set of 50\,sec collected by the Long Beach nodal array in southern California which contains 5200 sensors.
The top view of the sensor array is shown in Figure \ref{fig:map_view}.
Prior to applying the \acronym\ scheme, no reliable location estimation can be given by picked arrival times due to a large number of false picks as shown in Figure~\ref{fig:surface_fit}.
After applying \acronym\, the best-fitted hyperboloid surface from 3-D \acronym\ is shown as the red surface.
On a laptop, it takes just 31\,sec to finish this classification process, which is sufficient for real-time processing (note that the recording duration is 50\,sec).
At this point, we can use the \acronym\ picked arrival times to locate an event, but we must use a homogeneous medium assumption since there is no velocity model known prior to this experiment.
Based on the true picks classified by \acronym, this seismic event is recognized as a surface event whose location is shown by its epicenter marked by the red star in Figure \ref{fig:map_view}.
In order to verify our result, we schematically show the corresponding raw-data snapshot on the sensor array in Figure \ref{fig:sensorSnapshot}.
The gray-scale of the dots indicates the clipped signal amplitude on the corresponding sensor.
The red circle in Figure \ref{fig:sensorSnapshot} confirms that in the inverted time and location using the classified true picks, there is indeed a weak event that is barely visible in the array.
Moreover, the work log shows that there is a surface source in the estimated area but the local earthquake catalog has no record of earthquakes during the event time.

\subsection{Parameter selection}
Although it may seem that there are many parameters to be set for the \acronym\ method, they are actually tied to just one parameter that can be estimated from the data itself: $f_\text{dom}$, the dominant frequency of the source wavelet.
All parameters used in the above simulation examples are summarized in Table~\ref{tab:parameter_selection} and their recommended relationship to $f_\text{dom}$ is listed as well.
Cut-off frequency is chosen as the Nyquist frequency for $f_\text{dom}$.
Half period or wavelength is commonly used in signal detection and thus is chosen as the STW and ThresDist.
We follow the usual convention and choose LTW $= 10\,\times\,$STW.
Zero-crossing rate measures the long-term effects of the signal so we choose the same window length as LTW.
Gaussian smoothing tries to remove the short-term extreme values in the signal so we choose the same window as STW, $0.5T_\text{dom}$, which makes the smoothing window exactly the full wave period.
The choice of additive noise sigma being $T_\text{dom}$ makes the picked arrival time with noise lie in the waveform period with over 90\% probability (two sigmas of the normal distribution).
These parameters work well with the Ricker wavelet based simulations since the dominant frequency is a valid measurement of wavelet length (dominant period $T_\text{dom} \propto f_\text{dom}$).
However, sometimes the true wavelet length is longer than the dominant period ($T_\text{dom}$), e.g., a seismic trace with dispersion.
In this case, we can fix the cut-off frequency at 20\,Hz but make $T_\text{dom}$ longer to alleviate the problem.

\begin{table}
  \centering
  \caption{Parameters used in all simulations.}
  \label{tab:parameter_selection}
  \begin{tabular}{|c|c|}
    \hline
    Parameter & Recommended selection\\\hline
    Dominant frequency ($f_\text{dom}$) &	 10\,Hz\\\hline
    Dominant period ($T_\text{dom}$)	 &	  1/$f_\text{dom}$ sec\\\hline
    Cut-off frequency in LPF			 &	  $2 f_\text{dom}$\\\hline
    Short-term window (STW)				&	 $0.5 T_\text{dom}$ \\\hline
    Long-term window (LTW)				   &	$5 T_\text{dom}$\\\hline
    Zero-crossing rate window length (T)	 &	  LTW\\\hline
    Gaussian smooth function sigma		  &    STW\\\hline
    Threshold Distant (ThresDist)		 &	   $0.5 T_\text{dom}$ \\\hline
    RANSAC additive noise sigma			   &	ThresDist / 2\\\hline
  \end{tabular}
\end{table}

\section{Conclusion}

\label{sec:conclusion}

In this paper, we tackled the problem of phase association and event location estimation from arrival times by fitting a parametric model and then proposed an RANSAC-based fitting method (\acronym) to classify picked arrival times and detect possible events.
\acronym\ discriminates true event arrival times from false picks by associating them with some reasonable moveout curves.
Tests with synthetic data show that \acronym\ performs well for a 1-D linear array under layered medium assumption, as well for non-layered media and in the presence of dispersion.
\acronym\ is also expandable to the case of 2-D surface arrays by replacing the underlying hyperbolic curve model with a hyperboloid surface model.
The effectiveness of event location for the 2-D case is demonstrated in a 5200-element dense 2-D array for earthquake monitoring at Long Beach, CA.

In this study, we did not make any assumption on the seismic velocity in the study region.
Knowing the average velocity or a 1-D velocity model can be helpful to constrain the curvature of the fitted hyperbolas.
It is also possible to impose a constraint of P-S phase separation time to further improve the fidelity of the fitted curves.
This is beyond the scope of this paper and will be discussed in a follow-up work.

\section*{Acknowledgement}

\label{sec:acknowledge}
This work is supported by the Center for Energy and Geo Processing at Georgia Tech and King Fahd University of Petroleum and Minerals.
We are grateful to Zefeng Li for helpful discussions and the analysis of microearthquake data.
The seismic data analyzed in this study are owned by Signal Hill Petroleum, Inc. and acquired by NodalSeismic LLC.
We thank NodalSeismic LLC for making the one-week data available in this study.
LYC and ZP are partially supported by NSF award EAR-1818611.

\newpage
\bibliography{ransac}
\bibliographystyle{apalike}
\newpage
\appendix
\section{Parametric model of moveout curves}

\label{app:param_model}
In many microseismic applications, accurate velocity models may not be available.
However, a layered medium is commonly assumed for a shale rock region, in which case an estimate of the event location can be inferred from the arrival-time moveout curve across the monitoring geophone array.
Using arrival times not only has a clear physical meaning but also it turns out to be computationally efficient.
The primary requirement for this method to work is that there exists a parametric model $T(x)$ that approximates the arrival time $T$ versus horizontal offset $x$.
By estimating the finite number of model parameters, an event location can be uniquely determined.
Over the years, such parametric models have been gradually updated and generalized for various types of media.

\subsection{Homogeneous medium}
\begin{figure*}[hbt]
  \centering
  \begin{subfigure}[b]{0.45\textwidth}
    \centering
    \includegraphics[width=\textwidth]{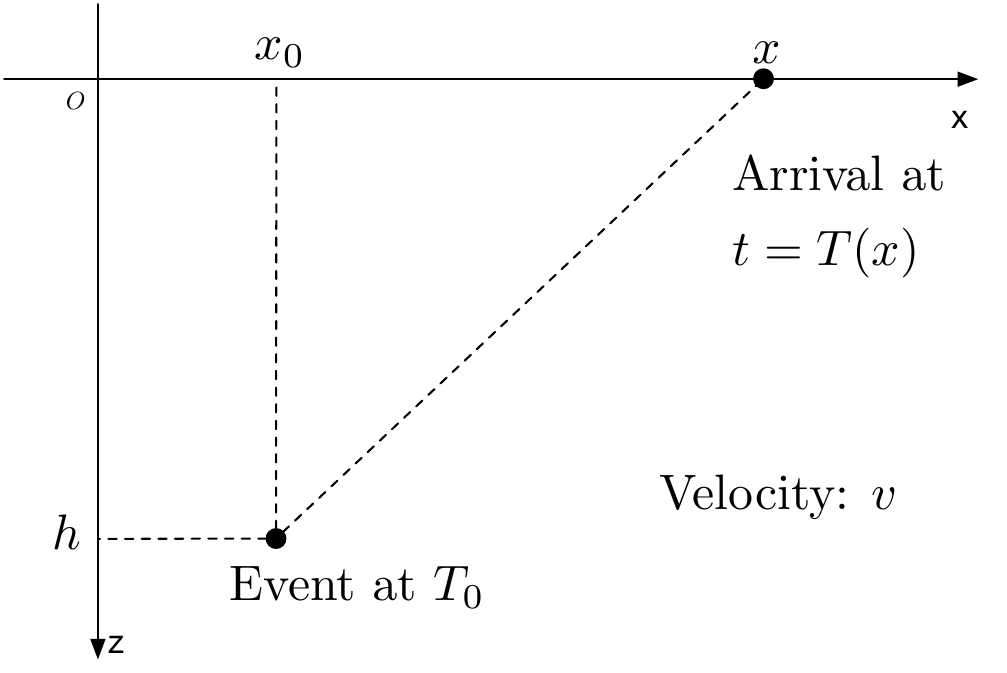}
    \caption[]%
    {{\small }}
    \label{fig:home_demo}
  \end{subfigure}
  \hfill
  \begin{subfigure}[b]{0.45\textwidth}
    \centering
    \includegraphics[width=\textwidth]{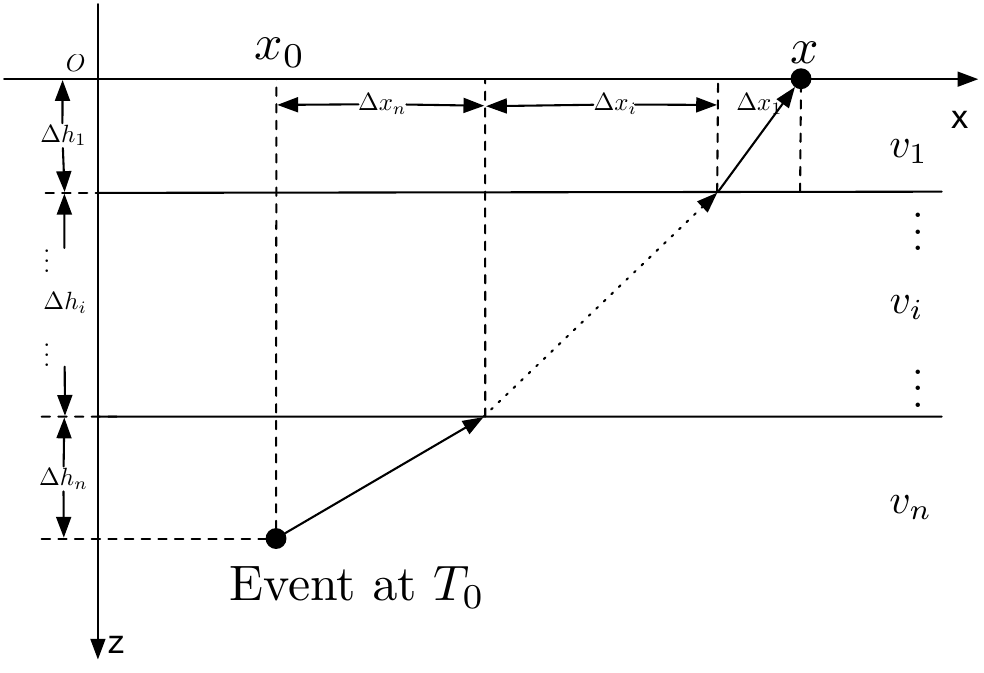}
    \caption[]%
    {{\small }}
    \label{fig:layered_demo}
  \end{subfigure}
  \caption[]
  {\small Geometry of ray paths and travel time for (a) homogeneous medium with velocity $v$, and (b) a layered media}
  \label{fig:geometry_media}
\end{figure*}

For a homogeneous medium, the geometry of ray tracing is shown in Figure~\ref{fig:geometry_media}a.
For an event originating at time $T_0$ and location $(x_0, h)$, a sensor at $x$ will receive the signal at time $T(x)$, so the relation between source-to-receiver travel time $T(x)-T_0$ and the horizontal offset $(x-x_0)$ is
\begin{equation}
  v^2 [T(x)-T_0]^2 = \underbrace{v^2 [T(x_0)-T_0]^2}_{=h^2} + (x-x_0)^2.
  \label{eq:TT_homo_raw}
\end{equation}
where we note that the zero-offset travel time is $T(x_0)-T_0 = h/v$.
The travel-time equation \eqref{eq:TT_homo_raw} can be rewritten in a form that is recognizable as the standard form of a hyperbola
\begin{equation}
  \frac{[T(x)-T_0]^2}{[T(x_0)-T_0]^2} - \frac{(x-x_0)^2}{v^2 [T(x_0)-T_0]^2} = 1.
  \label{eq:TT_homo_std}
\end{equation}
Thus, an event originating at time $T_0$ and location $(x_0, v[T(x_0)-T_0])$ can be uniquely determined by estimating the parameters $T_0$, $x_0$, $T(x_0)$, and $v$ in equation \eqref{eq:TT_homo_raw}, or \eqref{eq:TT_homo_std}.
The estimation involves fitting a hyperbola to the picked arrival times $T(x_j)$ in a linear surface array.

\subsection{Layered isotropic media}
A layered isotropic media, shown schematically in Figure~\ref{fig:geometry_media}b, is a little more complex than a homogeneous medium.
The travel time $\Delta T_i$ and horizontal offset $\Delta x_i$ of the $i$-th layer can be modeled as follows:
\begin{equation}
  \begin{aligned}
    \Delta x_i &= h_i \tan\theta_i = \frac{ph_iv_i}{\sqrt{1-(pv_i)^2}},\\
    \Delta T_i &= \frac{\Delta h_i}{v_i\cos\theta_i} = \frac{h_i}{v_i\sqrt{1-(pv_i)^2}},
  \end{aligned}
  \label{eq:TT_layered_single}
\end{equation}
where $p = \sin\theta_i / v_i$ is the ray parameter in Snell's law which is constant over all layers.
Within each layer, the travel time has a hyperbolic relationship with offset, i.e., $\Delta T_i(x_i)$ defines a hyperbola, so the overall travel time $\Delta T(x) = T(x) - T_0$ computed as the sum is not exactly a hyperbola
\begin{equation}
  \begin{aligned}
    \Delta x &= \sum_{i=1}^n \Delta x_i = \sum_{i=1}^n \frac{ph_iv_i}{\sqrt{1-(pv_i)^2}},\\
    \Delta T &= \sum_{i=1}^n \Delta T_i = \sum_{i=1}^n \frac{\Delta h_i}{v_i\sqrt{1-(pv_i)^2}}.
  \end{aligned}
\end{equation}
However, \citep{Dix1955} proved that a layered isotropic media behaves approximately like the homogeneous model when the offset $x$ is close to zero.
In other words, a hyperbolic moveout curve is observed near $\Delta x=0$ for an isotropic layered model with an equivalent velocity of
\begin{equation}
  v_\text{RMS}^2 = \frac{\sum\limits_{i=1}^n v_i^2 \Delta T_i}{\sum\limits_{i=1}^n \Delta T_i}.
  \label{eq:v_rms}
\end{equation}
In microseismic monitoring the receiving array is usually positioned over the top of the monitored events, so the offset $x$ should be close to zero and the approximation \eqref{eq:v_rms} can be used.
Moreover, \citet{Dix1955} gave a correction for a tilted layered model as well---the (approximate) relationship between $T$ and $x$ is still described by a hyperbolic curve
\begin{equation}
  \Delta T(x)^2 = \Delta T(0)^2+ \frac{\Delta x^2}{(v_\text{RMS}/\cos\theta)^2}.
  \label{eq:tilted_media}
\end{equation}
where $\theta$ is the tilt angle.

\subsection{Parametric model for TI media}
It is the parametric model rather than a hyperbolic curve that is essential to the data fitting method we will propose.
In cases of transverse isotropic (TI) media, \citet{Dellinger1993} gave an elliptic approximation of the arrival-time moveout curve
\begin{equation}
  \Delta T(x)^2= \frac{T(0)^4+(F_W+1)T(0)^2V_\text{NMO}^{-2}\Delta x^2+F_W^2V_\text{NMO}^4\Delta x^4}{T(0)^2+F_W^2V_\text{NMO}^{-2}\Delta x^2},
  \label{eq:TI_media}
\end{equation}
where $x$ is the offset, $T(0)$ is the vertical travel time, $V_\text{NMO}$ is the near-offset NMO velocity, and $F_W$ is a dimensionless anisotropy parameter.
Although equation \eqref{eq:TI_media} is a rational form with a fourth-order numerator and seems to be far from a hyperbolic curve, the basic idea is still valid: fitting a parametric model for $T(x)$ to locate an event.
In this paper, we use the simpler hyperbolic model as a demonstration.
The method to be proposed can be extended easily to other types of parametric models that adapt to different types of media.

\end{document}